\documentclass{emulateapj}
\usepackage[colorlinks,urlcolor=blue,citecolor=blue,linkcolor=blue]{hyperref} 
\usepackage{macros,graphicx,natbib}
\begin{document}

\title{The RRAT Trap: Interferometric Localization of Radio Pulses from J0628+0909}

\author{Casey J. Law\altaffilmark{1}}

\author{Geoffrey C. Bower}

\affil{Department of Astronomy and Radio Astronomy Lab, University of California, Berkeley, CA}

\author{Martin Pokorny}

\author{Michael P. Rupen}

\author{Ken Sowinski}
\affil{National Radio Astronomy Observatory, Socorro, NM}

\altaffiltext{1}{VLA Resident}

\keywords{pulsars: individual (J0628+0909), stars: neutron, techniques: interferometric}

\begin{abstract}
We present the first blind interferometric detection and imaging of a millisecond radio transient with an observation of transient pulsar J0628+0909. We developed a special observing mode of the Karl G.\ Jansky Very Large Array (VLA) to produce correlated data products (i.e., visibilities and images) on a time scale of 10 ms. Correlated data effectively produce thousands of beams on the sky that can localize sources anywhere over a wide field of view. We used this new observing mode to find and image pulses from the rotating radio transient (RRAT) J0628+0909, improving its localization by two orders of magnitude. Since the location of the RRAT was only approximately known when first observed, we searched for transients using a wide-field detection algorithm based on the bispectrum, an interferometric closure quantity. Over 16 minutes of observing, this algorithm detected one transient offset roughly 1\arcmin\ from its nominal location; this allowed us to image the RRAT to localize it with an accuracy of $1\dasec6$. With \emph{a priori} knowledge of the RRAT location, a traditional beamforming search of the same data found two, lower significance pulses. The refined RRAT position excludes all potential multiwavelength counterparts, limiting its optical luminosity to $L_{i'}<1.1\times10^{31}$\ erg s$^{-1}$ and excluding its association with a young, luminous neutron star.
\end{abstract}

\section{Introduction}
The fast ($\lesssim1$\ s) radio transient radio sky is populated by a wide range of source classes and physical processes. Neutron star pulses can be as brief as a nanosecond, making them the brightest objects in the universe and driving our understanding of coherent emission \citep{2003Natur.422..141H}. Observations at fast time scales are sensitive to pulsars and magnetars \citep{2000ApJ...541..367C}, extragalactic transients \citep{2003ApJ...596..982M,2007Sci...318..777L}, stars \citep{2008ApJ...674.1078O}, and planets \citep{1999JGR...10414025F}. Propagation effects, like dispersion and scattering, are also detectable at GHz frequencies on millisecond time scales and can be used to probe the interstellar and intergalactic media \citep{2002astro.ph..7156C}. 

A new class of fast transient is the rotating radio transient \citep[RRAT;][]{2006Natur.439..817M}. Like pulsars, RRATs are rotating neutron stars that emit radio pulses. Unlike pulsars, RRATs pulse rarely (roughly once out of every thousand rotations), so they are only detectable by their individual pulses. That observational distinction suggests that RRATs are physically distinct from ordinary pulsars \citep[although that is not yet clear;][]{2011MNRAS.415.3065K}. Identifying optical and X-ray counterparts to radio-detected RRATs would allow us to study the parent neutron star and define their relation to ordinary pulsars \citep{2009ApJ...702..692K}. Of the few dozen known RRATs \citep{2009ApJ...703.2259D,2010MNRAS.401.1057K,2010MNRAS.402..855B}, only one has a clear association with an X-ray counterpart, suggestive of a magnetar-like object \citep{2007ApJ...670.1307M,2009MNRAS.400.1439L}. However, since RRATs are an observationally defined class, it may be heterogeneous. Many more multiwavelength associations are needed before we can understand their underlying physics.

Large, single-dish telescopes have pioneered the study of RRATs and other fast radio transients through their high sensitivity and relatively simple signal processing requirements. Unfortunately, their design also limits their ability to survey efficiently, since survey speed scales as $N_{\rm{a}}^2 \, D^2$, where $N_{\rm{a}}$\ is the number of antennas used, and $D$\ is their diameter \citep{2008ASPC..395..225C,2011ApJ...742...12L}. More fundamentally, the arcminute-scale resolution of a large single-dish telescope is too coarse to find unique multiwavelength counterparts on arcsecond scales \citep{2009MNRAS.400.1445K}. Pulsar timing can obtain very precise localizations, but requires months or years of monitoring. Single-dish telescopes are also susceptible to radio frequency interference (RFI), compromising some of their most exciting potential discoveries \citep{2007Sci...318..777L,2010ApJ...715..939M}.

Interferometers, being composed of a distributed set of antennas, can overcome these limitations. They simultaneously have a large field of view and arcsecond resolution. Since interferometers are distributed over a large area, they are less susceptible to RFI and can localize interfering sources. However, pushing interferometric approaches down to subsecond time scales has seemed impractical due to the massive data throughput and computation needed to support image-based analysis. Novel signal processing concepts, such as the ``8gr8'' mode at the Westerbork Synthesis Radio Telescope \citep{2009A&A...498..223J}, ``V-FASTR'' \citep{2011ApJ...735...97W}, ``Fly's Eye'' \citep{2012ApJ...744..109S}, or pulsar gating \citep{2000ApJ...541..959B}, have opened access to more of the information available to interferometers. However, these concepts either require a priori information or are not based on visibilities, so they they lose some ability to blindly survey and localize transients.

Here we describe the first practical application of very high time-resolution observations with an interferometer, recording VLA visibilities every 10 ms, and using closure quantities to search for radio transients. Our goal was to improve the localization precision of RRAT J0628+0909 from arcminutes to arcseconds to facilitate a search for multiwavelength counterparts. Discovery observations with the Arecibo Observatory measured a pulse width of 10 ms, peak brightness of 85 mJy, and localization uncertainty of $3\damin5$ \citep{2006ApJ...637..446C,2009ApJ...703.2259D}, showing that it is accessible to our new VLA observing mode. Our observation marks the first use of the full VLA at this time scale and the first interferometric detection and localization of a fast radio transient. This also provides the first blind demonstration of a transient detection algorithm based on interferometric closure quantities \citep{2012ApJ...749..143L} and a second example of how to implement a general-purpose, fast correlator \citep{2011ApJ...742...12L}. We use the arcsecond localization of the transient to exclude its association with any cataloged multiwavelength counterpart, which excludes its association with a luminous, young, isolated neutron star.

\section{Observations}

\subsection{Developing a New Observing Mode}
Our observations required developing a nonstandard observing mode for the Wideband Interferometric Digital ARchitecture (WIDAR), the digital system at the heart of the VLA \citep{849012,2011ApJ...739L...1P}. To produce visibilities, the correlator architecture uses an ``XF'' design, where the cross-correlation of antenna voltages is done in custom FPGA hardware and the Fourier transform stage is performed in a commodity compute cluster known as the ``Correlator Backend''. Through a series of tests in early 2012, we tested the data throughput of WIDAR with a focus on correlator configurations that could use the full array on time scales much faster than 1 s.

In normal observing, the integration time out of the correlator is made as long as possible to minimize data volumes. The length of the integration is constrained by the requirement that the baselines not change significantly during an integration; that the atmosphere and the instrument remain stable; and that the visibilities can be flagged to remove RFI and other transitory effects, without removing too much useful astronomical data. For the VLA, this normally means dump times of $\sim1$\ second, which for a typical observing mode (cross-correlations for 27 antennas, two [128 MHz] subbands, and all four polarization products) gave total data rates of 1.4 MB/s. To detect individual pulses from RRATs like J0628+0909 requires dumping the data 100 times faster (10 ms), which in turn required removing several bottlenecks in the correlator system. The biggest speed improvement came from changes in the Correlator Backend software, which processes the visibilities and writes them out to disk; in particular, changes to the implementation of metadata and a reduction in the thread count. These and other improvements allow WIDAR (in some constrained correlator configurations) to achieve data rates as high as 140 MB/s.


\subsection{J0628+0909}
On 23 April 2012, we used the new mode to observe J0628+0909. Observations used three subbands with 25 antennas and two polarization products. For our 10 ms integration time, the total data rate was 90 MB s$^{-1}$. The three, 128-MHz subbands had central frequencies of 1312, 1440, and 1568 MHz. The highest of these was strongly affected by RFI and was ignored in this analysis.

We observed J0628+0909 for a total on-source time of 16 minutes. We observed J0632+103 as a complex gain calibrator and 3C 147 as a absolute flux density calibrator\footnote{Using the NRAO ``Perley-Butler 2010'' flux calibration standard.}. All observations used an integration time of 10 ms. The total data volume for the observation was about 200 GB.

\begin{deluxetable}{lc}
\tablecaption{Observing Configuration \label{obs}}
\startdata
\tablehead{
\colhead{Parameter} & \colhead{Value} \\ \hline
Antennas & 25, 25 m diameter \\
Configuration & C \\
Longest baseline & 12700 $\lambda \approx 3000$\ m\\
Channels per subband & 64 \\
Subband bandwidth & 128 MHz \\
Central frequencies & 1312, 1440, 1568 MHz \\
Polarizations & 2 \\
Integration time & 10 ms \\
Data rate & 90 MB s$^{-1}$ \\ \hline
\multicolumn{2}{c}{Target Field} \\ \hline
Direction (J2000) & (6h28m33s, +9\sdeg9\arcmin00\arcsec) \\
Duration & 16 minutes 
}
\enddata
\end{deluxetable}

After the observation, we discovered that the written data set was not complete. A random set of visibility spectra (all channels for a given subband, baseline, and integration) were written to disk as 0. The only trend seemed to be that the frequency of appearance of zeros increased with time to a maximum of roughly 50\% of all visibility spectra. This trend suggested that the correlator was not able to write data to disk as fast as our observing configuration required, despite our earlier success. We flagged these dropped data, which reduced our sensitivity by roughly 30\%.

\section{Analysis}
\label{analysis}

\subsection{Flagging and Calibration}
The data were calibrated using the Common Astronomy Software Package (CASA) version 3.3. Data identified as bad by the VLA on-line system were flagged, in addition to visibilities that were exactly equal to zero. Two channels with persistent RFI were flagged, but no time-dependent RFI flagging was done, since it is difficult to distinguish between RFI and astronomical transients. Since the bispectrum transient detection algorithm can distinguish RFI and transients, RFI flagging is helpful but not required.

As described below, our transient detection algorithm is based on the bispectrum, which is immune to gain phase errors. Gain amplitude calibration is needed, since the algorithm uses the relative values of all bispectra to measure spatial structure of transients. For a similar reason, we assume that all baselines have similar sensitivity, which is generally true for VLA observations at these frequencies; in the future, data weighting can be used to remove this assumption. Bandpass calibration is needed in order to coherently average the data in frequency. As such, the primary goal of calibration is to correct for bandpass, gain amplitude, and flux scale. 

In our case, the complex gain calibrator data were not useful due to human (CJL) error. We instead used the flux calibrator, 3C 147, for all calibration. Gain solutions were well behaved and images of the calibrated data toward 3C 147 showed a point source with the expected flux density. Since 3C 147 is more than 40\sdeg\ away from our target, it is not an ideal gain calibrator. It is reasonable to expect differences in gain phase relative to the target field due to their different paths of propagation. While gain amplitude calibration should transfer reliably between 3C 147 and our target field, imaging and astrometry are not guaranteed. We discuss these issues in detail in \S \ref{bispdet}.

\subsection{Transients Search}
One of the biggest challenges of this project was to efficiently search a large data volume for transients. Our strategy was to use a simple, efficient algorithm to identify candidate transients that were followed up in a more detailed analysis. For a first pass, we used a transient detection algorithm based on the bispectrum, the product of three visibilities from baselines that form a closed triangle\footnote{The phase of the bispectrum is another well-known closure quantity: the closure phase.} \citep{1987A&A...180..269C,1989AJ.....98.1112K,1995AJ....109.1391R}. As described in \citet{2012ApJ...749..143L}, the bispectrum is effective for transient detection because it is computationally efficient, sensitive to sources over a wide field of view, and immune to antenna-based gain phase errors. The standard deviation of bispectra is also a simple way of quantifying whether a source is point-like (as expected for fast transients) or extended; this functions as an efficient RFI rejection criterion. 

After calibration, all analysis was done in a custom Python radio interferometry data analysis package called tpipe\footnote{See \url{http://github.com/caseyjlaw/tpipe}.}. tpipe defines pipelines for analysis and visualization of visibilities with a focus on finding transient sources. The package uses radio interferometry libraries CASA (\url{http://casa.nrao.edu}), miriad-python \citep{2012PASP..124..624W}, and aipy \citep{2009AJ....138..219P}. Its design is flexible enough to read in Measurement Set or Miriad format data for either integration-based or dispersion-based transients searches.

Since the dispersion measure (DM) of J0628+0909 is roughly 88 pc cm$^{-3}$ \citep{2009ApJ...703.2259D}, we used a dispersion-based pipeline for our transients search. Across our lowest-frequency band, this DM introduced a time delay of roughly 40 ms, or four times our integration time. Failing to correct for that delay would have dramatically reduced our sensitivity. So we used a dispersion-based search that dedispersed visibilities before forming bispectra, images, or a phased-array beam. A dispersion track is defined for a given time, DM, and pulse width. In all analysis presented here, we assume a pulse width equal to the integration time of 10 ms, as seen in previous detections \citep{2009ApJ...703.2259D}.


The algorithm searched for transients by differencing visibilities in time. It started by dedispersing visibilities for all 300 baselines in our 25-antenna data set. Each set of dedispersed visibilities then had a mean background subtracted in time. The background was measured for each baseline and channel over a sliding window of 20 ms on either side of the dispersion track. The total background duration used was 40 ms, but this was spread over a somewhat larger window in time due to dispersion. To discern spectral structure in candidate pulses, each search was repeated for assumed DMs of 44, 88, and 132 pc cm$^{-3}$. 

After forming the background-subtracted, dedispersed visibilities, a bispectrum was calculated for all closed triangles in the array. For $n_a$ antennas, there are $n_a \, (n_a-1) \, (n_a-2)/6$ bispectra, all of which contain independent information \citep{1989AJ.....98.1112K}. For our observation, we calculated 2300 bispectra for each integration, DM, and polarization. After forming the bispectra, we took the mean over both polarizations to effectively produce a Stokes I bispectrum per integration and DM.

Because the visibilities were averaged in frequency before forming the bispectrum, beam smearing (chromatic aberration) decreases the sensitivity to sources far from the field center. The FWHM of the resulting delay beam is $\theta_{\rm{delay}} \approx 4 \, (\nu/\rm{BW}) \, \theta_{sb}$, where $\nu$\ is the observing frequency, BW is the bandwidth used in forming the bispectrum, and $\theta_{sb}$\ is the synthesized beam size \citep{tms,1999ASPC..180.....T}. Our search used a bandwidth of 256 MHz and a beam size of roughly 15\arcsec, producing a delay beam of 5\arcmin. This is large enough to cover the localization uncertainty of $3\damin5$ for J0628+0909 \citep{2009ApJ...703.2259D}, so the algorithm could be applied as described above\footnote{In cases where the delay beam is too small, one could form multiple delay beams or form the bispectrum from an average of frequency segments that have a larger delay beam. In the latter case, for $N$\ frequency segments, the noise per segment scales as $N^{1/2}$, but the SNR for the bispectrum scales as the SNR cubed, or $N^{-3/2}$ \citep[at low SNR;][]{1995AJ....109.1391R}. Averaging bispectra over segments benefits from another factor of $N^{1/2}$\, so overall the bispectrum loses sensitivity linearly with the number of segments used.}.

We searched for astronomical transients using both the real part of the mean and the standard deviation of the bispectra at each time and DM. The real part of the mean bispectrum detects any anomalous event, since it is proportional to the cube of the brightness of an event (both transients and RFI). The standard deviation over all bispectra in the array grows roughly as the square of the event brightness, if it is a point source. RFI has a different effect on the standard deviation of the bispectra, since it tends to be near the telescope or affect each baseline differently. The net effect is that RFI looks like a spatially extended transient, which has a large standard deviation over bispectra \citep{2012ApJ...749..143L}. So to select transients consistent with a celestial source, we required (1) a 5$\sigma$\ deviation of the mean bispectrum and (2) a standard deviation equal or less than that expected for a point source. For this and other datasets, this approach produced a list of candidate events that can be easily followed up manually.

Our first step in following up any candidate event was to look for associated RFI. Some RFI may have point-like spatial structure (e.g., from satellites), but such RFI almost always has narrow ($\lesssim 5$\ MHz) spectral structure. We searched for RFI by creating a spectrogram for times near the candidate. We did this by summing cross correlations to effectively form a phased-array beam at the phase center, assuming a phase-calibrated array. The vast majority of RFI was evident in a spectrogram like this. A second test of any candidate event was to image the dispersion trial and look for a point source. Finally, we moved the phase center to the location of any candidate source seen in an image, formed a spectrogram, and looked for the characteristic quadratic sweep of a dispersed pulse.

\section{Results}

\subsection{Event Detected with Bispectrum}
\label{bispdet}
\begin{figure}[tb]
\centering
\includegraphics[width=0.5\textwidth]{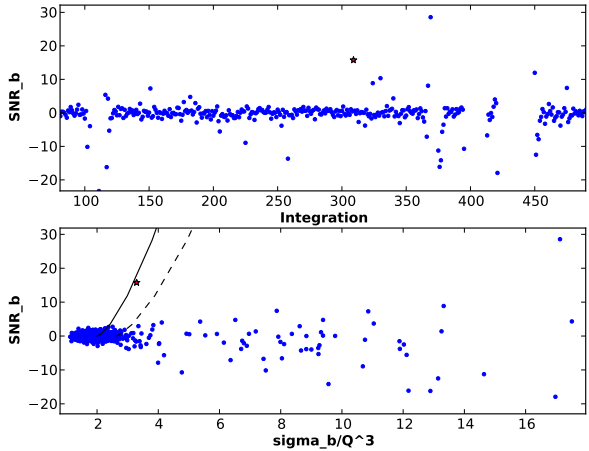}
\caption{(\emph{Top:}) Bispectrum light curve for roughly 4 seconds of 1 subband of VLA data toward J0628+0909. The red star shows the pulse detected by the algorithm. (\emph{Bottom:}) Mean versus the standard deviation of bispectra across the array for each integration. The standard deviation is normalized by an estimate of the noise per baseline cubed. The solid line shows the theoretical prediction for a point source with a range of brightnesses; a single, transient point source will appear on that line. The dashed line is the same as the solid line, but shifted to the right by 20\%. To tolerate some variance due to noise, we consider any point left of the dashed line to be a candidate. Both panels are scaled to show the transient, so RFI lies far above (in the top panel) or to the right (in the bottom panel).
\label{bisp}}
\end{figure}

We used the bispectrum transient detection algorithm to search 16 minutes of VLA data with two, 128-MHz subbands toward J0628+0909. With 10 ms integrations and three trial DMs, we searched roughly $2.8\times10^6$\ dispersion trials and found 14 candidate events. Of those candidates, all but one were traced to RFI or otherwise corrupted data. Figure \ref{bisp} summarizes the bispectrum algorithm during the candidate event when using only the lowest subband. It is detected by the bispectrum with roughly 16$\sigma$ significance\footnote{For convenience, we quote the apparent SNR measured in the mean bispectrum. In this case, the mean bispectrum is wider than a Gaussian, so the significance is overstated by a few percent.} and has a standard deviation consistent with the theoretical prediction for a point source.

For the candidate transient not clearly associated with RFI, we ran a series of tests to confirm the signal and show that it was associated with J0628+0909. First, the bispectrum search was repeated for data from the first and second subband independently. Similarly for polarization, we ran the algorithm for the RR and LL correlated products independently. The transient candidate was detected in all of these cases, showing that it is spectrally broad and not strongly circularly polarized.

\begin{figure}[tb]
\centering
\includegraphics[width=0.5\textwidth]{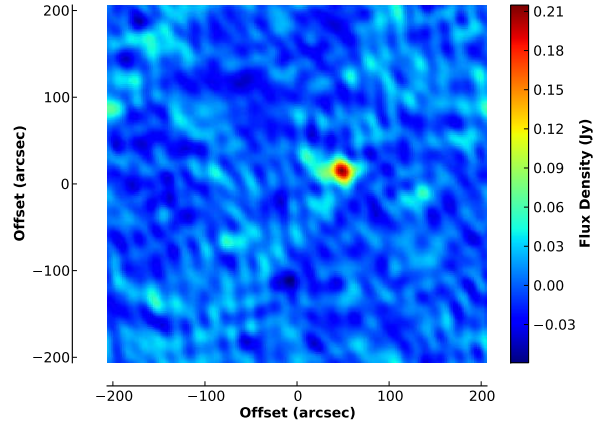}
\caption{Image of the dispersion trial with the candidate transient detected by the bispectrum algorithm and shown in Figure \ref{bisp}. The image has a beam size of 21\arcsec$\times$15\arcsec; the colorbar shows the flux scale in units of Janskys. A point source of significance 13$\sigma$\ is seen within 1\arcmin\ of the previous estimate for the location of RRAT J0628+0909.
\label{image}}
\end{figure}

Next, we imaged the background-subtracted dispersion trial associated with the candidate to confirm that it is point-like, as indicated by the standard deviation of the bispectra. Figure \ref{image} shows an image of the first subband of the VLA data for this candidate. The image contains a single point source with brightness of $215\pm17$\ mJy beam$^{-1}$, or $13\sigma$\ significance. 

For an observation with 25 antennas, an event detected with apparent $16\sigma$\ significance in the bispectrum should have roughly $15\sigma$ significance in the image plane, assuming good phase calibration \citep{2012ApJ...749..143L}. The slight reduction in the significance of detection in the image plane is consistent with the fact that the image noise is 20\% higher than expected (considering dropped data). These two points suggest that the gain calibration was good to roughly 20\%, consistent with the reasonable image quality. We note that imaging with two subbands shows imaging artifacts indicative of stronger phase errors, so we focus our subsequent analysis on the lowest subband, covering frequencies from 1248 to 1376 MHz.

We fit an elliptical Gaussian representative of the synthesized beam to the point source to measure its location and uncertainty. The point source is offset from the J2000 phase center of (6h28m33s, +9\sdeg09\arcmin00\arcsec) by ($48\dasec1$, $14\dasec9$), which places it at (6d28m36.25s, 9\sdeg9\arcmin14$\dasec$9). This offset is within the localization uncertainty of previous detections ($3\damin5$), which clearly associates it with J0628+0909. It is also within the delay beam of our search (5\arcmin), which shows that it is the same pulse seen by the bispectrum algorithm. The uncertainty in the Gaussian centroid in (RA, Dec) is (1$\dasec$3, 1$\dasec$5).

To test the astrometric accuracy of the data, we measured location of a steady source elsewhere in the field. We imaged NVSSJ062859+091149 in a series of 10 1-s snapshot images around the time of the bright pulse from J0628+0909. The cataloged source location is (6:28:59.99, +9:11:49.7) known to a precision of roughly 1\arcsec. The mean position measured in our snapshot images is (6:28:59.93, +9:11:49.5) with a standard deviation of (1$\dasec$0, 0$\dasec$6) and confirms its astrometric accuracy. Adding this uncertainty in quadrature to the source centroid error, we find an absolute centroid precision of (1$\dasec$7, 1$\dasec$6).

\begin{figure}[tb]
\centering
\includegraphics[width=0.5\textwidth]{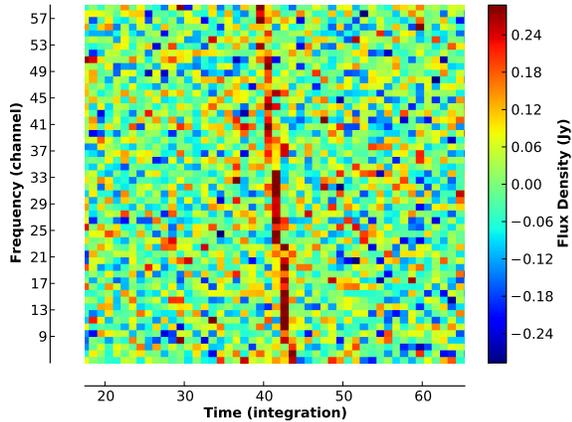}
\caption{Spectrogram showing the dispersed pulse from RRAT J0628+0909 shown in Figure \ref{bisp}. The spectrogram is generated for a phased-array beam pointed at the transient seen in the image in Figure \ref{image}. The color scale is in units of Janskys.
\label{spec}}
\end{figure}

As final confirmation of the detection, we formed a beam on the measured location of the pulse and generated a spectrogram. Figure \ref{spec} shows the spectrogram of the first subband toward the transient. The transient is clearly a broad-band pulse with an unresolved width of 10 ms. The pulse spans five integrations with the characteristic quadratic sweep expected of dispersion.

\begin{figure}[tb]
\centering
\includegraphics[width=0.5\textwidth,bb = 0 170 600 600,clip]{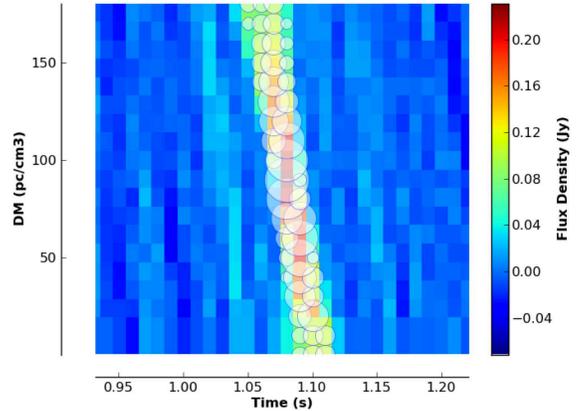}
\caption{Pulse SNR seen in phased-array data for a range of dispersion trials toward RRAT J0628+0909. Circles mark dispersion trials with SNR greater than 5 and scale linearly with SNR.
\label{disp}}
\end{figure}

Figure \ref{disp} shows the apparent pulse brightness in a phased-array beam as a function of assumed DM and pulse time. This kind of analysis mimics traditional, single-dish pulse searches, which rely on the dispersive delay to distinguish an astronomical source from RFI. We find a peak pulse brightness for DM near 90 pc cm$^{-3}$, consistent with previous detections.

\subsection{Events Detected with Phased Array}
With \emph{a priori} constraints on the location of J0628+0909, we searched for pulses with a phased-array, dispersion-based search. We expected to detect more pulses with a phased-array beam than with the bispectrum, since the bispectrum sensitivity falls off as the cube of source brightness for SNR per baseline less than 1 \citep{1995AJ....109.1391R}. For our single VLA subband data with 10 ms integrations, this happens for a brightness of roughly 280 mJy beam$^{-1}$.

Since the data are largely unflagged, a phased-array beam is highly susceptible to RFI. We filtered candidate transients by finding events with peak flux for an assumed DM near 88 pc cm$^{-3}$. As with candidates from the bispectrum search, we rejected candidates associated with RFI seen in spectrograms. This search revealed two further candidate pulses with peak phased-array beam SNR of 6 and 7. Imaging dispersion trials at their peak SNR shows point sources at the same location as the pulse detected with the bispectrum algorithm. All told, our VLA observations find three pulses with brightnesses of 75, 86, 215 mJy beam$^{-1}$\ in 10 ms. These flux densities may be supressed by up to 20\% due to phase calibration errors.

With three pulse detections, we can use the relative pulse arrival times to estimate the underlying rotation period of the RRAT. Assuming that each pulse arrives from a random rotation of the neutron star, the period is at most the greatest common divisor of their relative arrival times \citep{2006Natur.439..817M}. We find that a period of 1.24148 s is most consistent with the rotation period of the neutron star, confirming earlier work.

\subsection{Using the Improved Localization}
Our arcsecond localization of J0628+0909 makes it easier to search for counterparts. No compelling counterpart was found in a search with the Simbad, NED, and HEASARC systems. A search of X-ray data archives did not find any \emph{Chandra} observation that could potentially detect thermal emission from the neutron star.

\begin{figure}[tb]
\centering
\includegraphics[width=0.5\textwidth,bb = 0 120 640 670,clip]{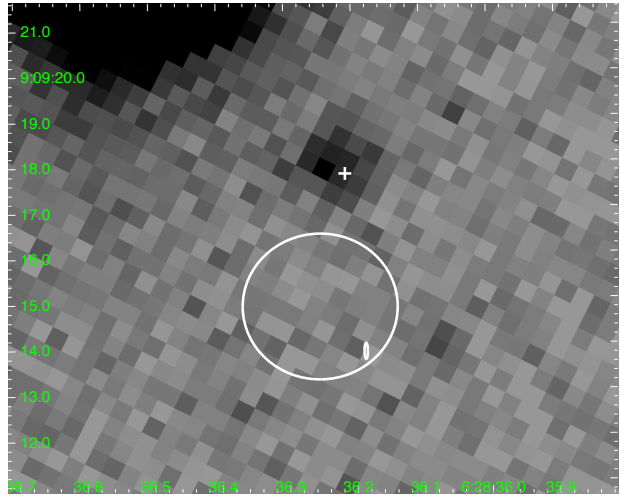}
\caption{SDSS i'-band image of the region around J0628+0909. The large ellipse marks the localization uncertainty for the brightest radio pulse we detect from J0628+0909. This includes uncertainty in the absolute astrometry described in \S \ref{bispdet}. The plus shows the location of SDSSJ062836.22+090917.5, the nearest SDSS counterpart. The small ellipse shows a new localization of J0628+0909 from an observing campaign with the Arecibo Observatory (Nice et al 2012, in prep).
\label{cpart}}
\end{figure}

A search of the Sloan Digital Sky Survey \citep[SDSS-III, data release 9;][]{2000AJ....120.1579Y} found one potential association for the RRAT. Figure \ref{cpart} shows a SDSS-III i'-band image with the large ellipse showing the 1$\sigma$\ uncertainty on our best-fit location for J0628+0909. A red stellar source, SDSSJ062836.22+090917.5, is offset by 2$\sigma$\ from the radio source centroid. We use the offset of these two sources and the density of sources brighter than SDSSJ062836.22+090917.5 to estimate the probability of random association \citep[e.g.,][]{2011ApJ...740...87C}. For a i'-band source density of 4.4$\times10^{-3}$\ and a radio source offset of 3$\dasec$5, we expect a 15\% chance of false association, suggesting that the association is not significant. 

Independent of our VLA observation, the team that discovered J0628+0909 has been monitoring it (Nice et al 2012, in prep). In Arecibo observations spanning over three years, they have collected thousands of pulses that allow a precise localization of the RRAT using traditional pulsar timing techniques \citep{2006MNRAS.369..655H}. Using these observations, they have localized J0628+0909 to (6:28:36.1832$\pm$0.0025, +9:9:13.93$\pm$0.18). As shown in Figure \ref{cpart}, the timing solution is within the 1$\sigma$\ uncertainty in our localization, confirming the image-based analysis.

By excluding an SDSS counterpart to J0628+0909, we can put an upper limit on its optical luminosity. The typical limiting magnitude is i'$=21.3$\ mag, which corresponds to an apparent flux density limit of $1.1\times10^{-5}$\ Jy \citep{1996AJ....111.1748F}. The dispersion of pulses toward J0628+0909 is 88 pc cm$^{-3}$, implying a distance of 2.5 kpc \citep[typically better than 30\% accurate;][]{2002astro.ph..7156C}. At that distance, we expect extinctions of $A_V=0.98$\ mag and $A_{i'}=0.62$\ mag \citep{2003A&A...409..205D,1989ApJ...345..245C}. Together, these effects imply a upper limit on the i' luminosity of $1.1\times10^{31}$\ erg s$^{-1}$. It is difficult to compare this to the optical/infrared properties of isolated neutron stars, since observations --- and the spectral and temporal behavior of neutron stars themselves --- are heterogeneous \citep{2010MNRAS.407.1887R}. However, the SDSS i' limit does exclude luminous, young neutron stars \citep[e.g., the Crab;][]{1997ApJ...477..465E}, which have flat infrared spectra and can be orders of magnitude brighter than the SDSS limit.

\section{Conclusion}
We have demonstrated a new observing mode for the VLA that can write 10 millisecond visibilities, opening access to this exciting timescale to image and visibility analysis techniques. We used this new observing mode to find and image individual pulses from the poorly-localized RRAT J0628+0909. This is the first blind demonstration of our pulse detection algorithm based on the interferometric closure quantity known as the bispectrum and the first use of visibilities to detect and localize a fast radio transient.

We measure a new position for J0628+0909 of (6d28m36.25s, 9\sdeg9\arcmin14$\dasec$9) (J2000) with an accuracy of roughly 1$\dasec$6. Our interferometric localization is two orders of magnitude better than the localization made in its discovery observation. It is also consistent with new, sub-arcsecond localization made via a traditional timing analysis with Arecibo data spanning more than three years. Our localization for J0628+0909 excludes its association with any cataloged multiwavelength counterpart. SDSS imaging places an upper limit on the RRAT optical luminosity that excludes association with a young, luminous, isolated neutron star.

This paper presents the first real-world application of new signal processing and algorithms for using radio interferometers to the study millisecond transients. More than just pushing traditional slow transient techniques to fast time scales, our study of J0628+0909 shows how fast interferometric imaging encompasses and extends single-dish techniques. Future iterations can expand this concept by implementing real-time, all-time processing, opening the latent survey ability of radio interferometers.

\acknowledgements{We thank Steve Croft, Jason Hessels, and Adam Ginsburg for useful discussions and Julia Deneva and David Nice for sharing early results of their Arecibo observations. We are grateful to the DRAO team (particularly Brent Carlson, Peter Dewdney, David Fort, and Sonja Vrcic) for designing and producing the WIDAR correlator, whose capabilities made this work possible. 

This work was supported by NSF grant AST-0807151. The National Radio Astronomy Observatory is a facility of the National Science Foundation operated under cooperative agreement by Associated Universities, Inc.. Funding for SDSS-III has been provided by the Alfred P. Sloan Foundation, the Participating Institutions, the National Science Foundation, and the U.S. Department of Energy Office of Science. This research has made use of NASA's Astrophysics Data System.}

{\it Facility:} \facility{VLA}, \facility{Sloan}


\end{document}